**Ecohydrological land reanalysis**

Yohei Sawada[1], Hiroyuki Tsutsui[2], Hideyuki Fujii[3], and Toshio Koike[2]

[1] Institute of Engineering Innovation, the University of Tokyo, Tokyo, Japan

[2] International Centre for Water Hazard and Risk Management, Public Work Research Institute, Tsukuba, Japan

[3] Earth Observation Research Center, Japan Aerospace eXploration Agency, Tsukuba, Japan

Corresponding author: Y. Sawada, Institute of Engineering Innovation, the University of Tokyo, Tokyo, Japan, 2-11-6, Yayoi, Bunkyo-ku, Tokyo, Japan, yohei.sawada@sogo.t.u-tokyo.ac.jp

**Abstract**

The accurate estimation of terrestrial water and vegetation is a grand challenge in hydrometeorology. Many previous studies developed land data assimilation systems (LDASs) and provided global-scale land surface datasets by integrating numerical simulation and satellite data. However, vegetation dynamics has not been explicitly solved in these land reanalysis datasets. Here we present the newly developed land reanalysis dataset, ECoHydrological Land reAnalysis (ECHLA). ECHLA is generated by sequentially assimilating C- and X- band microwave brightness temperature satellite observations into a land surface model which can explicitly simulate the dynamic evolution of vegetation biomass. The ECHLA dataset provides semi-global soil moisture from surface to 1.95m depth, Leaf Area Index (LAI), and vegetation water content and is available from 2003 to 2010 and from 2013 to 2019. We assess the performance of ECHLA to estimate soil moisture and vegetation dynamics by comparing the ECHLA dataset with independent satellite and in-situ observation data. We found that our sequential update by data assimilation substantially improves the skill to reproduce the seasonal cycle of vegetation. Data assimilation also contributes to improving the skill to simulate soil moisture mainly in the shallow soil layers (0-0.15m depth). The ECHLA dataset will be publicly available and expected to contribute to understanding terrestrial ecohydrological cycles and water-related natural disasters such as drought.

**1. Introduction**

Accurate monitoring of soil moisture and terrestrial ecosystem is crucially important to quantify global water and carbon cycles (e.g., Oki and Kanae 2006; Cox et al. 2000), land-atmosphere interactions (e.g., Seneviratne et al. 2010), and water-related natural disasters such as drought events (e.g., Sheffield et al. 2012; Zhou et al. 2014). It is a grand challenge in hydrometeorology to provide accurate and spatiotemporally continuous datasets of land surface hydrological variables, which can be called land reanalysis datasets, by integrating land surface models (LSMs) and global-scale observations.

Several land reanalysis datasets have been developed. The pioneering work of Rodell et al. (2004) developed Global Land Data Assimilation System (GLDAS) to provide land surface hydrological data using multiple state-of-the-art LSMs driven by bias-corrected meteorological forcing data. Balsamo et al. (2015) developed the ERA-interim/land dataset based on the European Centre for Medium-Range Weather Forecasts (ECMWF) land surface model and the bias-corrected ERA-interim atmospheric reanalysis dataset. The Modern-Era Retrospective Analysis for Research and Applications version 2

(MERRA-2) used observation-based precipitation to force an LSM and provided the skillful soil moisture estimation (Reichle et al. 2017a). While these land reanalysis products did not sequentially adjust the soil moisture estimation from the LSMs by assimilating hydrological observations, the Soil Moisture Active Passive mission Level-4 surface and root-zone soil moisture (SMAP L4) product (Reichle et al. 2017b) realized sequential data assimilation of the SMAP microwave brightness temperature observation into an LSM (see also De Lannoy and Reichle 2016a, 2016b) and provided global surface and root-zone (0-1.0m) soil moisture data constrained by the satellite observation. Although many useful land reanalysis data have been provided, to our best knowledge, no reanalysis product explicitly simulated the interactions between vegetation growth and hydrological processes.

There are several Land Data Assimilation Systems (LDASs) which explicitly solve not only soil moisture but vegetation dynamics such as Leaf Area Index (LAI) by integrating LSMs and satellite observations. For example, LDAS-Monde allowed sequential data assimilation of satellite-observed surface soil moisture and LAI into an LSM which can explicitly solve the dynamic evolution of vegetation biomass (Albergel et al. 2017). This LDAS could improve the skill to simultaneously simulate surface soil moisture and vegetation dynamics (Albergel et al. 2018). Kumar et al. (2019) and Kumar et al. (2020) developed the LDAS to assimilate satellite-observed LAI and microwave Vegetation Optical Depth (VOD) data, respectively. Sawada and Koike (2014) and Sawada et al. (2015) developed the Coupled Land and Vegetation Data Assimilation System (CLVDAS) in which satellite microwave brightness temperature observations were directly assimilated into an LSM with a dynamic vegetation model. CLVDAS successfully improved the skill to simultaneously simulate surface and root-zone soil moisture and LAI and has been applied to drought monitoring and forecasting (Sawada and Koike 2016; Sawada 2018; Sawada et al. 2019).

We aim to develop the new land reanalysis dataset, called ECoHydrological Land reAnalysis (ECHLA), using CLVDAS and the long-term C- and X-band microwave brightness temperature observations from Advanced Microwave Scanning Radiometer for Earth observation system (AMSR-E) and AMSR2. The main objective of this paper is to assess the quality of ECHLA.

## 2. Method
### 2.1. Overview of CLVDAS
Figure 1 shows the schematic of CLVDAS. Our offline LSM can simulate both soil moisture and vegetation water content using meteorological forcing data. Simulated soil moisture, vegetation water content, and surface temperature were used to estimate C- and X-band microwave brightness temperature by our Radiative Transfer Model (RTM). Based on our estimated brightness temperature and observed brightness temperature from AMSR-E and AMSR2, the LSM's state variables were sequentially updated by the particle filter data assimilation algorithm. This assimilation cycle was repeated to generate the spatio-temporally continuous soil moisture and vegetation water content dataset. Although original CLVDAS has a model parameter optimization module (Sawada and Koike 2014; Sawada 2020), we did not use it to provide the ECHLA dataset. In this paper, the brief description of CLVDAS and the modification from the previous version are presented. See Sawada and Koike (2014) and Sawada et al. (2015) for the complete description of CLVDAS.

### 2.2. Land Surface Model: EcoHydro-SiB
Our LSM, EcoHydro-SiB, uses the one-dimensional Richards equation to solve vertical interlayer water flows (Wang et al. 2009). Capillary suction and hydraulic conductivity are estimated by the van Genuchten's water retention curve (van Genuchten 1980). Evaporation, transpiration, and the other heat fluxes are estimated by the parameterization of the Simple Biosphere version 2 (SiB2, Sellers et al. 1996).

Net primary production is also estimated by the SiB2 photosynthesis-conductance parameterization. The dynamic evolution of vegetation biomass is calculated considering carbon allocation, turnover, and stress loss. Both water and temperature stress factors are considered to calculate the carbon allocation fractions and stress loss using the parameterization of Ivanov et al. (2008). The water stress factor is calculated from the vertical profile of soil moisture considering the distribution of root biomass modelled by Jackson et al. (1996).

To connect the estimated carbon biomass to microwave brightness temperature, vegetation water content (VWC) should be obtained (see also the section 2.2). First, LAI is obtained by:

$LAI = S_{la} \times C_{leaf}$ (1)

where $S_{la}$ is the specific leaf area and $C_{leaf}$ is the carbon pool of leaves. Then, LAI is converted to VWC by the linearized version of the empirical equation proposed by Paloscia and Pampaloni (1988):

$$VWC = \frac{LAI}{y} \quad (2)$$

where y is a constant parameter. While the exponential relationship between LAI and VWC originally proposed by Paloscia and Pampaloni (1988) has been used in the previous version of CLVDAS (Sawada et al. 2015), here we used equation (2) since it is found that our data assimilation works more stably with equation (2) in the dense forested areas. The empirical parameter, y, was set to 1.5 based on the in-situ observation of Sawada et al. (2017). The similar linear relationship between VWC and optically observable vegetation indices has been proposed by the previous works (e.g., Paloscia et al. 2013; Gao et al. 2015; Santi et al. 2017).

## 2.2. Radiative Transfer Model

In our Radiative Transfer Model (RTM), the microwave radiative transfer on land surface and vegetation canopy is calculated by the tau-omega model (Mo et al. 1982) so that the emission from the land surface attenuated by the canopy, the emission from the canopy, and the emission from the canopy reflected by the land surface are considered. The emission and attenuation by the canopy are modelled by the single scattering albedo and VOD. We assumed the linear relationship between VOD and VWC following Jackson and Schmugge (1991). In the C- and X-band microwave, the relationship between VOD and VWC was found to be species-independent (Sawada et al. 2017) so that we can connect VWC estimated by EcoHydro-SiB (see equation (1-2)) to VOD without any species-dependent parameters.

Land surface emissivity in the microwave range is strongly controlled by surface soil moisture. In our RTM, the Advanced Integral Equation Model (AIEM) with the incorporation of a shadowing effect developed by Kuria et al. (2007) is used to calculate land surface emissivity from surface soil moisture estimated by EcoHydro-SiB.

Since satellite level microwave brightness temperature is estimated based on both surface soil moisture and VWC, it is expected that our data assimilation of the microwave brightness temperature observation positively impacts the simulation of both surface soil moisture and vegetation water content. In addition, our data assimilation also impacts the unobservable sub-surface soil moisture since it is strongly related to the observable surface soil moisture and vegetation dynamics (see also Sawada et al. 2015).

## 2.3. Data Assimilation: Particle Filter

The sampling-importance-resampling Particle Filter (PF) is used as the data assimilation method in CLVDAS. PF is a Monte Carlo approximation of the Bayes' theorem. The flavors of PF have been intensively used in the hydrometeorological applications (e.g., Moradkhani et al. 2005; Qin et al. 2009; Abbazadeh et al. 2018; Poterjoy et al. 2019). The advantage of PF is the flexibility to non-linear dynamics and non-Gaussian error distribution. The disadvantage of PF is the expensive computational cost. However, in the application of the one-dimensional LSM, we can perform PF with the reasonable computational cost.

From the initial conditions obtained in the previous data assimilation cycle, we perform the ensemble forecasting and evaluated each ensemble member (or "particle") by calculating the likelihood or the closeness to the observation. We use the Geman-McClure type estimator (Geman and McClure 1987) to evaluate the likelihood function since it is robust to the outlier of observations (Sawada et al. 2015). Based on the calculated likelihood, the weights of the particles are calculated. Then, the particles are resampled based on the weights by multinomial draws. The weight of the particle should be the probability that the particle is selected for the next data assimilation cycle, so that particles with larger weights are copied many times and particles with smaller weights are likely to be excluded in the resampling step.

After particles are resampled, perturbations should be added to the state variables of the resampled particles to maintain the diversity of the particles. Perturbing particles is important to prevent the filter degeneracy (e.g., Snyder et al. 2008), in which all but one particle have extremely small weights in the resampling step. In the previous version of CLVDAS, we added small perturbations to carbon pool of leaves and soil moisture of all soil layers by drawing from the bounded uniform distribution (Sawada et al. 2015). In this study, we draw the perturbation from the Gaussian distribution whose mean is the state variables of a resampled particle. The variance of the Gaussian distribution is set to be proportional to the variance in all prior particles following Moradkhani et al. (2005). In addition, the perturbations to each state variables are designed to follow the covariance of state variables of prior particles so that the perturbations are physically consistent. The similar strategy has been adopted in Kawabata and Ueno (2020).

## 3. Data

The ERA5 global atmospheric reanalysis dataset (Hersbach et al. 2020) was used to drive our LSM, EcoHydro-SiB. The meteorological forcings needed to drive EcoHydro-SiB were surface pressure, precipitation, surface air temperature, relative humidity, incoming solar radiation, incoming longwave radiation, and wind speed. For precipitation, incoming solar radiation, and incoming longwave radiation, we used the 10-member ensemble forecast to consider the uncertainty in meteorological forcing (see also section 4). The data were linearly interpolated to 0.25 degree and hourly.

As microwave brightness temperature observations, the AMSR-E and AMSR2 L3 products (Kachi et al. 2013) were used. Brightness temperature observations at 6.925GHz and 10.25GHz were used. Both horizontally and vertically polarized observations were used. We used only night scene data since the effect of surface temperature errors on the estimation of microwave brightness temperature by RTM is small at night (Liu et al. 2011). The native spatial resolution of the L3 products is 0.1 degree and the data were resampled to 0.25 degree by box averaging. The temporal resolution is approximately 2-daily.

The International Satellite Land Surface Climatology Project 2 (Global Soil Data Task Group 2000) and the Food and Agricultural Organization (FAO) global dataset (FAO 2003) were used to derive the model parameters of EcoHydro-SiB.

To evaluate the skill of ECHLA to simulate vegetation dynamics, the optical LAI product processed by Ichii et al. (2017) was used. Ichii et al. (2017) performed detailed quality control to MODerate resolution Imaging Spectroradiometer (MODIS) onboard Terra and Aqua satellite LAI L4 data (MCD15A2H; Myneni et al. 2015). The spatial and temporal resolutions are 0.25 degree and 8-daily, respectively.

To evaluate the skill of ECHLA to simulate surface soil moisture, the European Space Agency Climate Change Initiative Soil Moisture (ESA CCI SM) v05.2 (Dorigo et al. 2017) was used. We chose the combined passive and active microwave product. The horizontal and spatial resolution of ESA CCI SM is 0.25 degree and approximately 2-daily, respectively.

To evaluate the skill of ECHLA to simulate sub-surface soil moisture, the in-situ soil moisture observations archived at International Soil Moisture Network (ISMN) (Dorigo et al. 2011) were used. We used the data at the depths of up to 1.95m during our study period (i.e., 2003-2019; see section 4). The data whose observation period is less than 1-year were excluded in our analysis, and the data whose quality flags are not "G" (Good) were also excluded (Dorigo et al. 2013). In addition to the ISMN in-situ observation, we used the in-situ soil moisture observations of the Yanco flux tower site (34.99S, 146.29E) in Australia which is the JAXA's calibration and validation site for AMSR2 land products.

## 4. Experiment Design

The ECoHydrological Land reAnalysis (ECHLA) dataset was generated by driving CLVDAS. The timestep of the LSM, EcoHydro-SiB, was 1 hour. The spatial resolution of EcoHydro-SiB was 0.25 degree. EcoHydro-SiB had 20 soil layers from surface to 1.95m depth. The thickness of the first surface layer was set to 0.05m while that of the other soil layers was set to 0.10m. Since the sensed soil depth of AMSR-E and AMSR2 observations is shallow, the thickness of the surface soil layer was set to thinner than the other layer to effectively assimilate microwave brightness temperature. The data assimilation window was set to 1 day so that the LSM's state variables were updated whenever microwave brightness temperature observations could be obtained. The ensemble size of PF was set to 25. The 11 ensemble members of ERA5 (10-member ensembles and their ensemble mean) were randomly assigned to each ensemble member of EcoHydro-SiB. Our study area was limited to the global snow-free area. We excluded the pixels in which climatological snow-covered periods estimated by ERA5 data were longer than 15-day. This is because our current RTM does not consider the effects of snow on brightness temperature. Considering the effects of snow accumulation in our RTM may bring additional errors in the land reanalysis, which was avoided in this study.

Table 1 summarizes the specification of the ECHLA dataset. The spatial and temporal resolutions of the ECHLA dataset are 0.25 degree and daily, respectively. There are two data periods in ECHLA. The first period is from 2003 to 2010, when the data were generated by assimilating AMSR-E brightness temperature observations into EcoHydro-SiB. The second period is from 2013 to 2019, when AMSR2 brightness temperature observations were used for data assimilation. Since there are no full-year observations from AMSR-series satellites in 2011-2012, no ECHLA data are currently provided in this period. The ensemble mean of soil moisture and VWC will be publicly available.

We assessed the ECHLA dataset by comparing it with independent satellite and in-situ data described in section 3. Since there are few available and reliable data of VWC, we compared ECHLA-estimated LAI with MODIS LAI. We extracted the observation depth of each ISMN in-situ soil moisture observation and compared the in-situ soil moisture observation with ECHLA-estimated soil moisture at the corresponding depth. The evaluation metrics used in this study were Mean Absolute Error (MAE), unbiased Root-Mean-Square-Error (ubRMSE), and correlation coefficient (R), which have been widely used to evaluate the skill to estimate land surface variables such as soil moisture (e.g., Balsamo et al. 2015; Reichle et al. 2017a, 2017b; Albergel et al. 2018).

To quantify the benefit of data assimilation to estimate soil moisture and vegetation dynamics, we performed the Open Loop (OL) experiment in which EcoHydro-SiB is run without data assimilation. We compared the performance of ECHLA with that of the OL experiment.

## 5. Results

Figure 2 shows the performance of ECHLA to reproduce the vertically polarized 6.925 GHz brightness temperature observation of AMSR-E and AMSR2. Overall, MAE and ubRMSE are less than 2[K] and 6[K], respectively, and R is larger than 0.6 (Figures 2a, 2c, and 2e). MAE is relatively larger in the desert areas than the other areas (Figure 2a) because although the uncertainty in surface soil roughness parameters of RTM substantially affects the skill to simulate brightness temperature on bare soil (e.g., Kuria et al. 2007; Sawada et al. 2017), these surface soil roughness parameters were not intensively calibrated. R is relatively lower in the rain forest areas (Figure 2e) because there is the small seasonal variation of brightness temperature in the dense vegetated areas. Unphysical patterns of ubRMSE can be found in Egypt due probably to the strong effects of Radio Frequency Interference (RFI) (see Moesinger et al. 2020 for the distribution of the RFI effects). Data assimilation substantially improves the simulation of brightness temperature all over the world (Figures 2b, 2d, and 2f). Our data assimilation effectively constrains the land state variables by microwave brightness temperature observations. Similar improvements can be found in the assimilated brightness temperatures with the other frequencies and polarizations (not shown).

Figure 3 shows the performance of ECHLA to reproduce the MODIS LAI observation. Data assimilation greatly improves the performance of ECHLA to simulate LAI. Figures 3a and 3b show that data assimilation reduces MAE mainly in the densely vegetated areas such as Amazon, Congo, and southeast Asia. Although data assimilation has the relatively small impact on MAE in the arid and semiarid areas (Figure 3b), the MAE in these areas is already small in the OL experiment. On the other hand, data assimilation substantially reduces ubRMSE and increases R in the arid and semiarid areas (Figures 3c-f), which reveals that data assimilation of microwave brightness temperature greatly improves the skill to reproduce the seasonal cycles of LAI. The improvement of the LAI simulation is marginal in the United States, which may be attributed to the RFI effects on C-band microwave vegetation signals (Moesinger et al. 2020).

Figure 4 shows the performance of ECHLA to reproduce the surface soil moisture observation of the ESA CCI SM dataset. Data assimilation negatively impacts MAE (Figures 4a and 4b). The lowest soil moisture during drydown periods in ESA CCI SM is systematically higher than that in ECHLA and this bias cannot be corrected by our data assimilation. However, data assimilation improves ubRMSE and R mainly in the semi-arid sparsely vegetated areas and deserts (Figures 4c-f) so that the representation of seasonal cycles and anomalies of surface soil moisture is improved.

Figure 5 summarizes the performance of ECHLA to reproduce the surface and sub-surface soil moisture observation of the ISMN in-situ soil moisture dataset. We used 1327 observations in total. The results are stratified by 4 soil depths: 0-0.05m (n=171), 0.05-0.15m (n=387), 0.15-0.45m (n=348), and 0.45m-2.05m (n=421). The medians of MAE for 4 soil depths are 0.0765, 0.0746, 0.0784, and 0.0865, from surface to deep soil layers (Figure 5a). The medians of ubRMSE for 4 soil depths are 0.0601, 0.0551, 0.0468, and 0.0451, from surface to deep soil layers (Figure 5d). The medians of R for 4 soil layers are 0.643, 0.581, 0.530, and 0.306, from surface to deep soil layers (Figure 5g). Although the median of MAE at the shallow soil layers (0-0.05m and 0.05-0.15m) is improved, the impact of data assimilation on the skill to reproduce ISMN in-situ soil moisture observation is marginal (Figures 5b, 5e, and 5h).

Figures 6-9 indicate that many of in-situ observations sites in which the skill scores are degraded by data assimilation are located in the United States (see also Figures S1-S8 for the comparison between observation networks in the United States and Australia). When we excluded the data from the United States, MAE of the surface soil layer (0-0.05m) is improved in 75% of the observation sites and ubRMSE of the shallow soil layers (0-0.05m and 0.05-0.15m) is improved in 75% of the observation sites although the magnitude of the improvement is small (Figures 5c, 5f, and 5i). Figures 5c, 5f, and 5i also reveal that the number of observation sites in which data assimilation substantially degrade the skill to reproduce in-situ soil moisture observation is small outside the United

States. It implies that our data assimilation of C- and X-bands brightness temperatures into an LSM suffers from the relatively strong RFI effects in the United States (Moesinger et al. 2020).

Figure 10 shows the timeseries of LAI and soil moisture at the Yanco site which is the JAXA's calibration and validation site for AMSR2 land observations. Figure 10a clearly shows that ECHLA reproduces the seasonal cycles of LAI more accurately than OL. Table 2 shows ubRMSE and R are improved by data assimilation, which is consistent to our global-scale evaluation. Figures 9b-f show that data assimilation impacts soil moisture drydown behaviors. Table 2 shows that this impact of data assimilation slightly improves ubRMSE and R in the sub-surface soil moisture simulation.

**6. Discussion**

In this study, we propose the new land reanalysis, called ECoHydrological Land reAnalysis (ECHLA). We sequentially assimilate microwave brightness temperature and frequently adjust the LSM's simulation, which is the unique feature compared with the existing land reanalysis datasets generated by the free-run of LSMs such as ERA-interim/land and MERRA2. Compared with the other approaches which sequentially assimilate microwave observations such as SMAP L4, the unique feature of ECHLA is that the spatiotemporally continuous vegetation data are available. The spatiotemporally continuous water and vegetation data may contribute to monitoring and quantifying drought (e.g., Sawada et al. 2014; Asoka and Mishra 2015; Kumar et al. 2021). Our data assimilation of the brightness temperature observations significantly improves the simulation of vegetation dynamics, which was consistent to our previous works (Sawada et al. 2015; Sawada 2018; Sawada et al. 2019).

Our data assimilation also improves ubRMSE and R of surface soil moisture against ESA CCI and MAE and ubRMSE of surface and sub-surface soil moisture against the ISMN soil moisture at the depth of 0-0.15m although ECHLA suffers from the relatively strong RFI effects in the United States. The impact of data assimilation on deeper zone soil moisture is marginal. The in-situ soil moisture observations used in this study are mainly located in the relatively arid areas. In the arid and semi-arid grassland areas, water stress strongly affects vegetation dynamics so that our PF can inversely estimate root-zone soil moisture from the observation of aboveground vegetation growth (Sawada et al. 2015). However, grasses have more than 50% of their root in the top 15cm according to Jackson et al. (1996), which makes it difficult to infer soil moisture in deeper soil layers from the observation of aboveground vegetation dynamics. In addition, the coupling between surface and root-zone soil moistures is weak in the arid areas (Kumar et al. 2009), which makes it difficult to infer deeper-zone soil moisture dynamics from surface soil moisture observations.

In the ECHLA dataset, ubRMSE is lower than 0.06 in the majority of the in-situ observation sites in the United States and Australia, which is compatible to the skill of the similar products such as the SMAP L4 dataset (Reichle et al. 2017b). It should be noted that the study period, study area, in-situ observations used, and validation approach of our study are substantially different from SMAP L4 (Reichle et al. 2017b) so that it is difficult to directly compare the value of evaluation metrices of the two products. Data assimilation changes the behavior of soil moisture during drydown periods, which is also consistent to Reichle et al (2017b).

There are several limitations in ECHLA and we have several suggestions toward the next improved version of ECHLA. First, the current study area is limited to the snow-free regions due to the lack of a snow module of our RTM. By considering the impact of snow on radiative transfer in canopy (e.g., Tsutsui and Koike 2012), we will provide the global dataset. Second, our current data assimilation does not consider the uncertainty in the LSM's parameters, such as hydraulic conductivity. In the previous works, optimizing the unknown parameters by microwave observations substantially improved the skill of LSMs (e.g., Yang et al. 2007; Bandara et al. 2013, 2015; Sawada and Koike 2014). Recently, CLVDAS obtained an efficient parameter optimization and uncertainty quantification module which can be applied globally (Sawada 2020). The parameter optimization is expected to improve the skill to simulate soil moisture in the relatively deep zone soil layers since the value of parameters directly affects the whole model's dynamics (Sawada 2020). In the next version of ECHLA, optimized model parameters and their quantified uncertainties will be used and our limited improvement of the deep zone soil moisture estimation by data assimilation will expected to be overcome. Third, although we focused on the use of AMSR-series satellites since they have been providing long-term brightness temperature observations, the other satellites such as L-band observations by SMAP and Soil Moisture and Ocean Salinity (SMOS) may be beneficial to further improve ECHLA since we can improve the sensing depth and avoid the RFI effects. We will use multiple satellites to provide spatiotemporally continuous ecohydrological datasets.

## 7. Conclusions

In this paper, we propose the new land reanalysis, ECHLA, and provide the overview and validation of the product. A semi-global soil moisture (up to 1.95m depth) and VWC dataset from 2003 to 2010 and from 2013 to 2019 is provided. The land state variables are sequentially adjusted by assimilating C- and X- band brightness temperature observations into the LSM. We find that the data assimilation of the brightness temperature observation from AMSR-E and AMSR2 substantially improves the skill to simulate vegetation dynamics and moderately improves the skill to simulate surface and shallow root-zone soil moisture. The ECHLA dataset will be publicly available from the JAXA's G-Portal (https://gportal.jaxa.jp/gpr/?lang=en).

## Acknowledgements

This study was supported by JAXA grant ER2GWF102 and JSPS KAKENHI Grant 17K18352, 18H03800, and 21H01430. Computational resources were provided by Oakforest-PACS supercomputer at the University of Tokyo.

**Data Availability Statement**

The ISLSCP II dataset can be downloaded at https://daac.ornl.gov/cgi-bin/dataset_lister.pl?p=29. The FAO dataset can be downloaded http://www.fao.org/land-water/land/land-governance/land-resources-planning-toolbox/category/details/en/c/1026564/. The AMSR-E and AMSR2 dataset was provided by JAXA and can be downloaded at https://gportal.jaxa.jp/gpr/?lang=en (please search in this portal web site). The MODIS LAI dataset can be downloaded at ftp://modis.cr.chiba-u.ac.jp/ichii/DATA/MODIS/. The ESA CCI soil moisture dataset can be downloaded at https://www.esa-soilmoisture-cci.org/. The ISMN in-situ soil moisture dataset can be downloaded at https://ismn.geo.tuwien.ac.at/en/.

**Table 1.** Specification of the ECHLA dataset.

| Spatial resolution | 0.25 degree × 0.25 degree |
|---|---|
| Temporal resolution | daily |
| Period | 2003-2010, 2013-2019 |
| Variables | Volumetric soil moisture at the depth of up to 1.95m [$m^3/m^3$] |
| | Vegetation water content [$Kg/m^2$] |

**Table 2.** The skill score of ECHLA in the JAXA calibration and validation site at Yanco, Australia

|  | MAE | | ubRMSE | | R | |
|---|---|---|---|---|---|---|
|  | ECHLA | OL | ECHLA | OL | ECHLA | OL |
| LAI | 0.109 | 0.0623 | 0.467 | 0.662 | 0.639 | 0.465 |
| Soil moisture (3cm) | 0.000864 | 0.00138 | 0.0532 | 0.0525 | 0.852 | 0.853 |
| Soil moisture (10cm) | 0.00159 | 0.00101 | 0.0423 | 0.0435 | 0.886 | 0.871 |
| Soil moisture (15cm) | 0.0813 | 0.0737 | 0.0258 | 0.0265 | 0.844 | 0.833 |
| Soil moisture (45cm) | 0.0648 | 0.0556 | 0.0451 | 0.0458 | 0.612 | 0.599 |
| Soil moisture (75cm) | 0.0747 | 0.0681 | 0.0450 | 0.0493 | 0.475 | 0.271 |

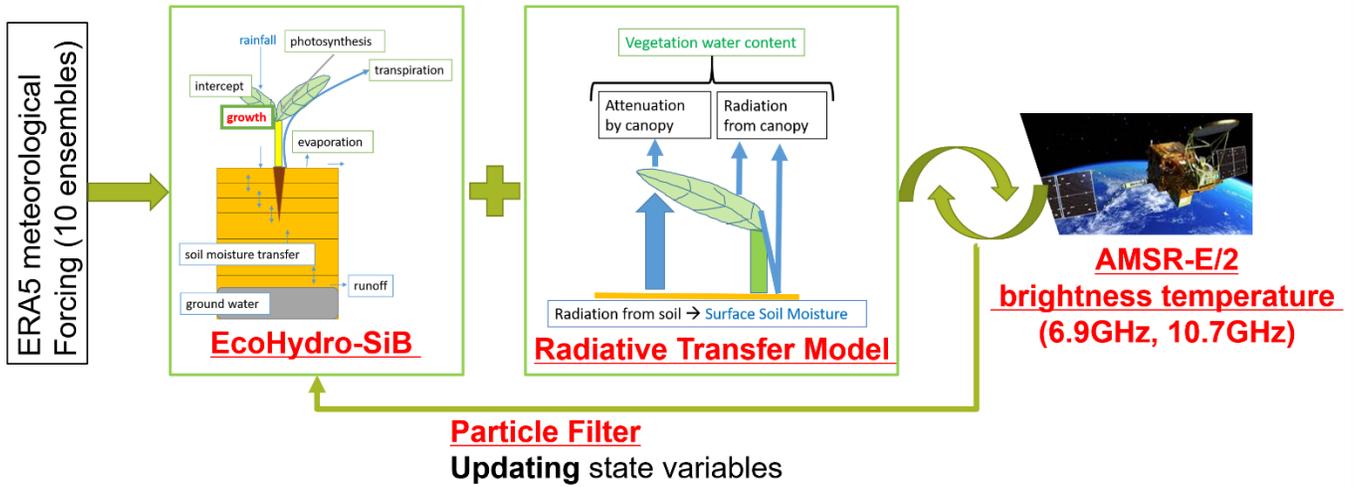

**Figure 1.** Schematic of CLVDAS. See section 2 for details.

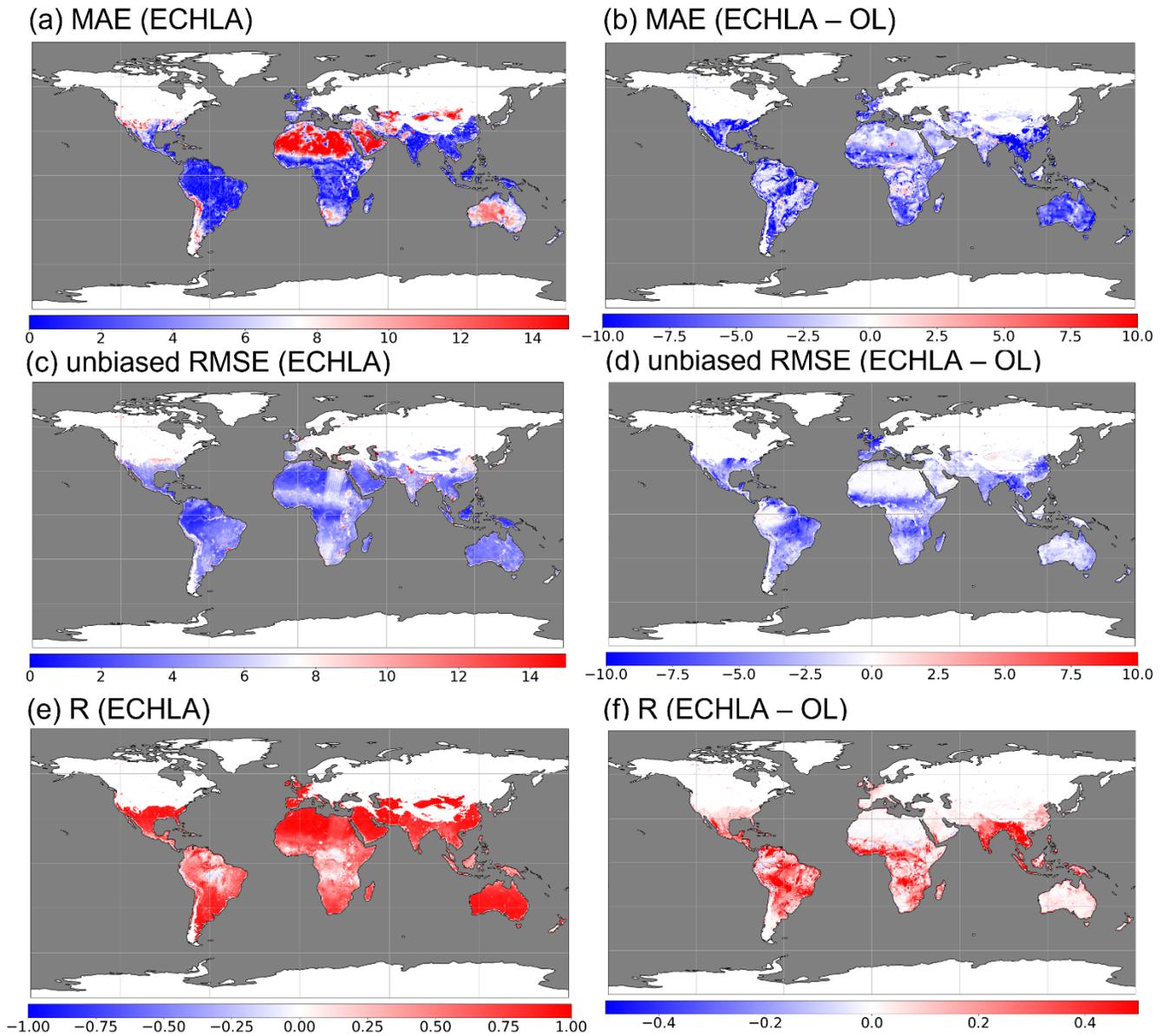

**Figure 2.** The performance of ECHLA to reproduce AMSR-E and AMSR2-observed vertically polarized brightness temperature at 6.925GHz. (a) MAE, (c) ubRMSE, and (e) R of simulated brightness temperature of ECHLA. The differences of (b) MAE, (d) ubRMSE, and (f) R between ECHLA and the OL experiment (see also section 4).

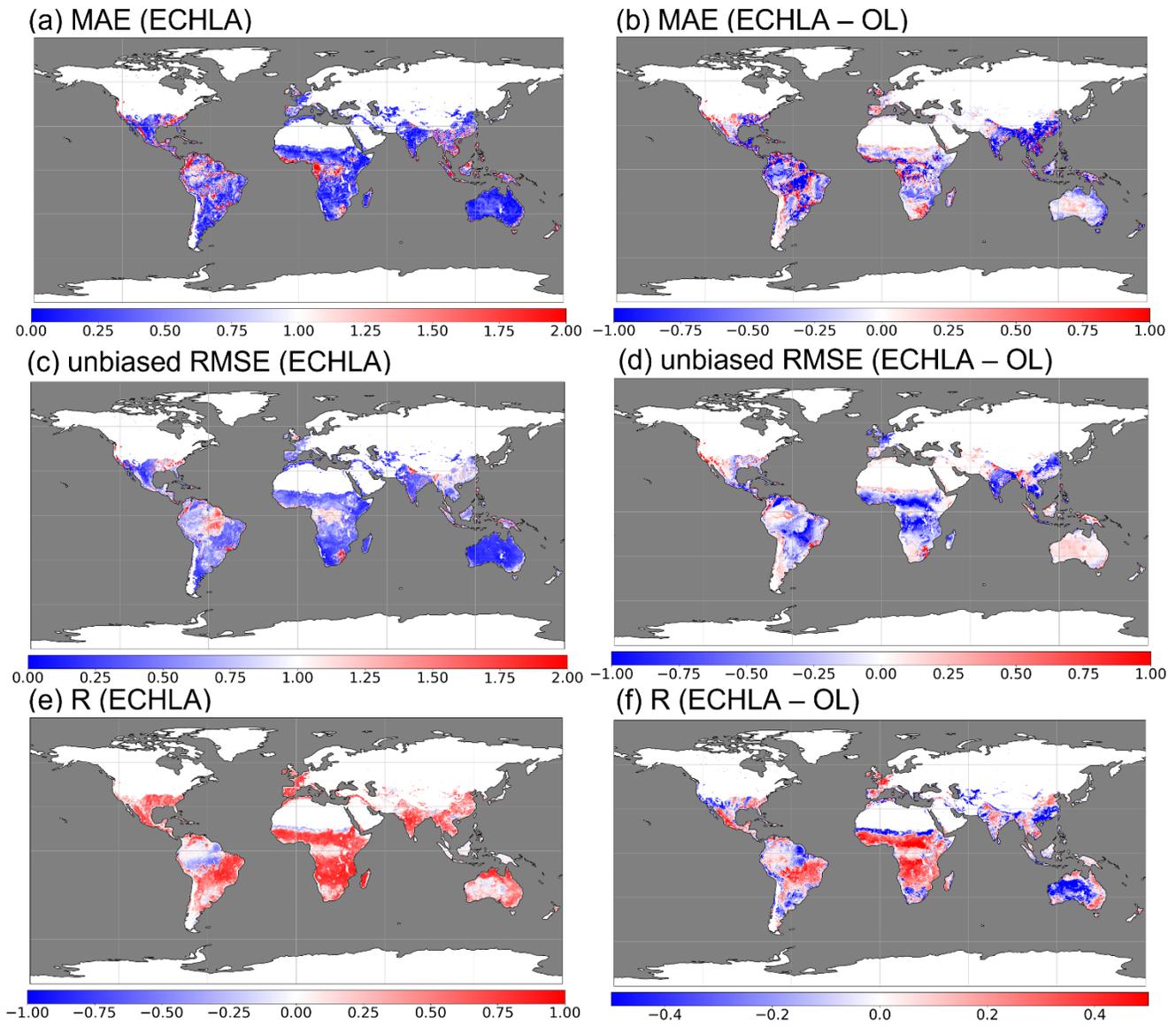

**Figure 3.** The performance of ECHLA to reproduce MODIS LAI. (a) MAE, (c) ubRMSE, and (e) R of simulated LAI of ECHLA. The differences of (b) MAE, (d) ubRMSE, and (f) R between ECHLA and the OL experiment (see also section 4).

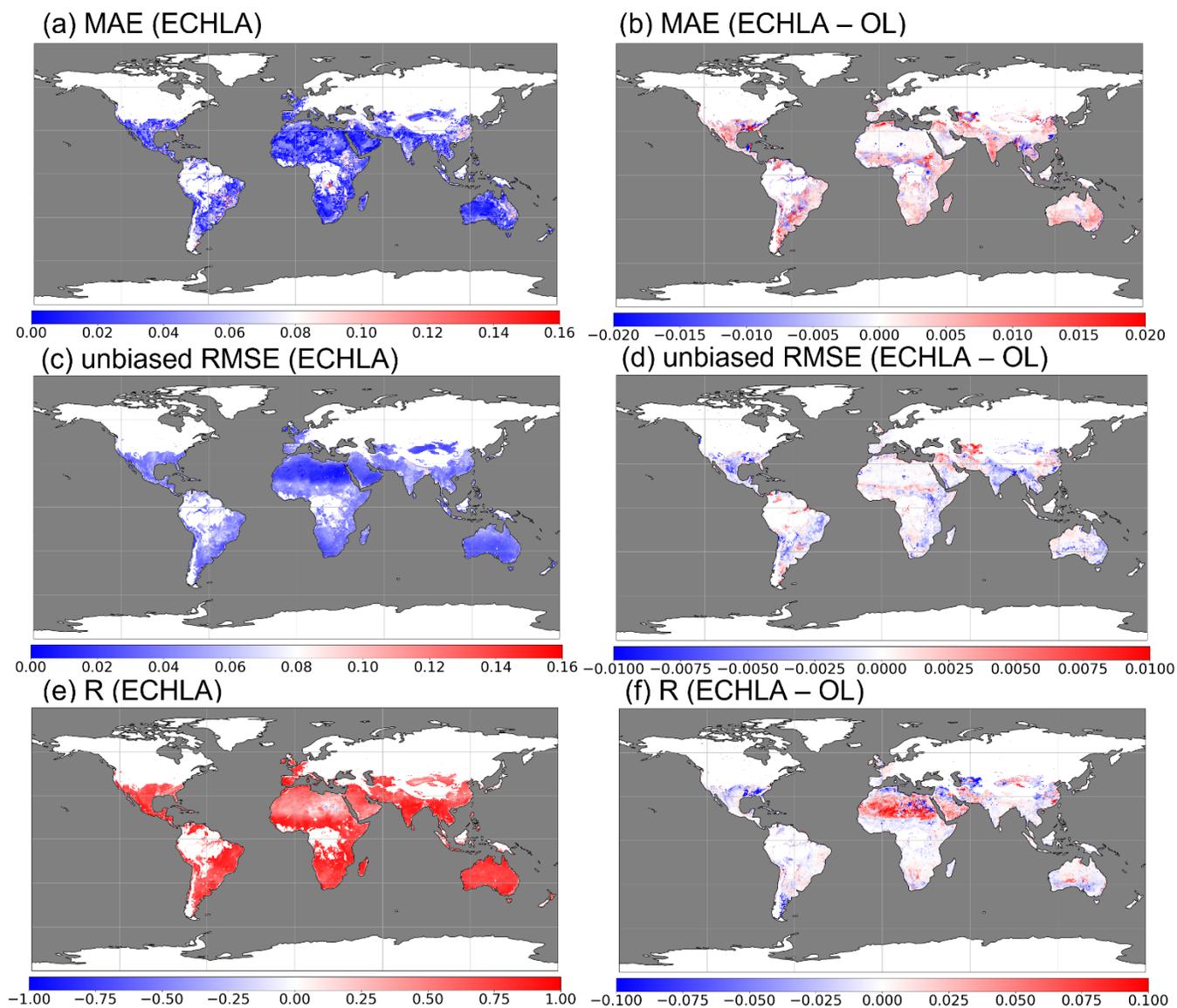

**Figure 4.** The performance of ECHLA to reproduce ESA CCI surface soil moisture. (a) MAE, (c) ubRMSE, and (e) R of simulated surface (0-0.05m depth) soil moisture of ECHLA. The differences of (b) MAE, (d) ubRMSE, and (f) R between ECHLA and the OL experiment (see also section 4).

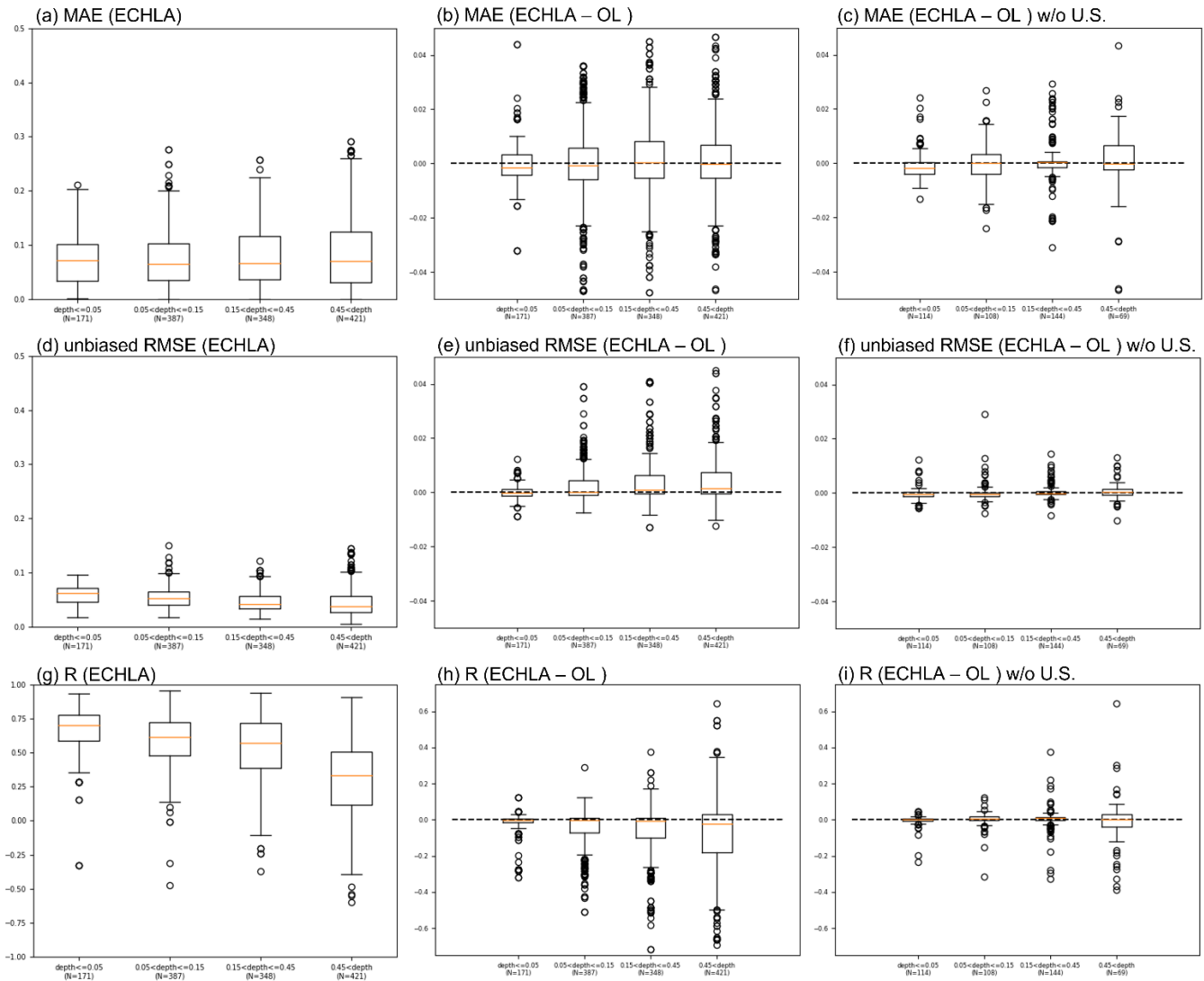

**Figure 5.** The performance of ECHLA to reproduce ISMN in-situ soil moisture observations. (a) MAE, (d) ubRMSE, and (g) R of simulated soil moisture of ECHLA. The differences of (b) MAE, (e) ubRMSE, and (h) R between ECHLA and the OL experiment (see also section 4). (c,f,i) same as (b,e,h) but without in-situ soil moisture observation in the United States

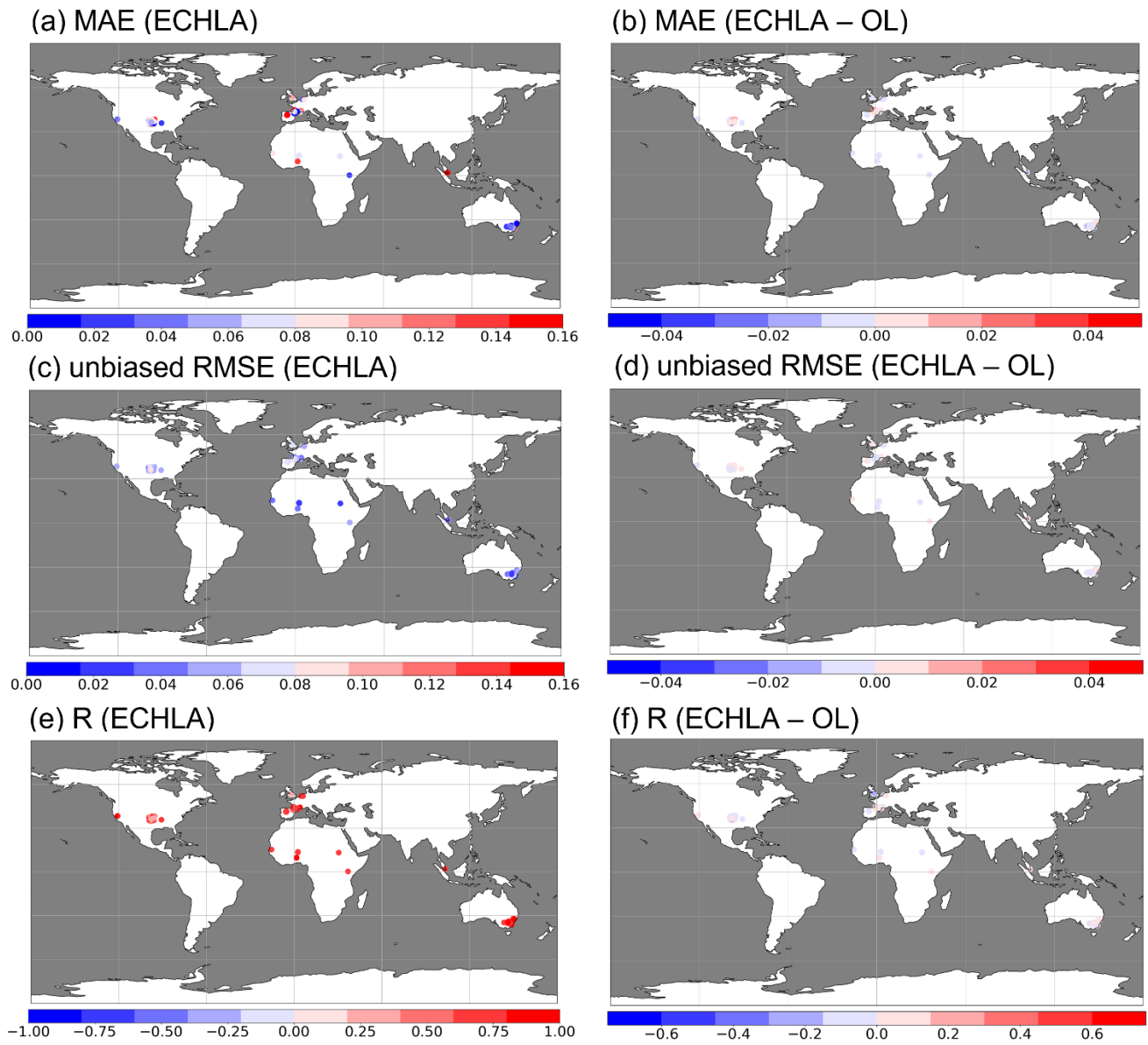

**Figure 6.** Spatial distribution of (a) MAE, (c) ubRMSE, and (e) R of ECHLA soil moisture against ISMN in-situ soil moisture at the depth of 0-0.05m. The differences of (b) MAE, (d) ubRMSE, and (f) R between ECHLA and the OL experiment.

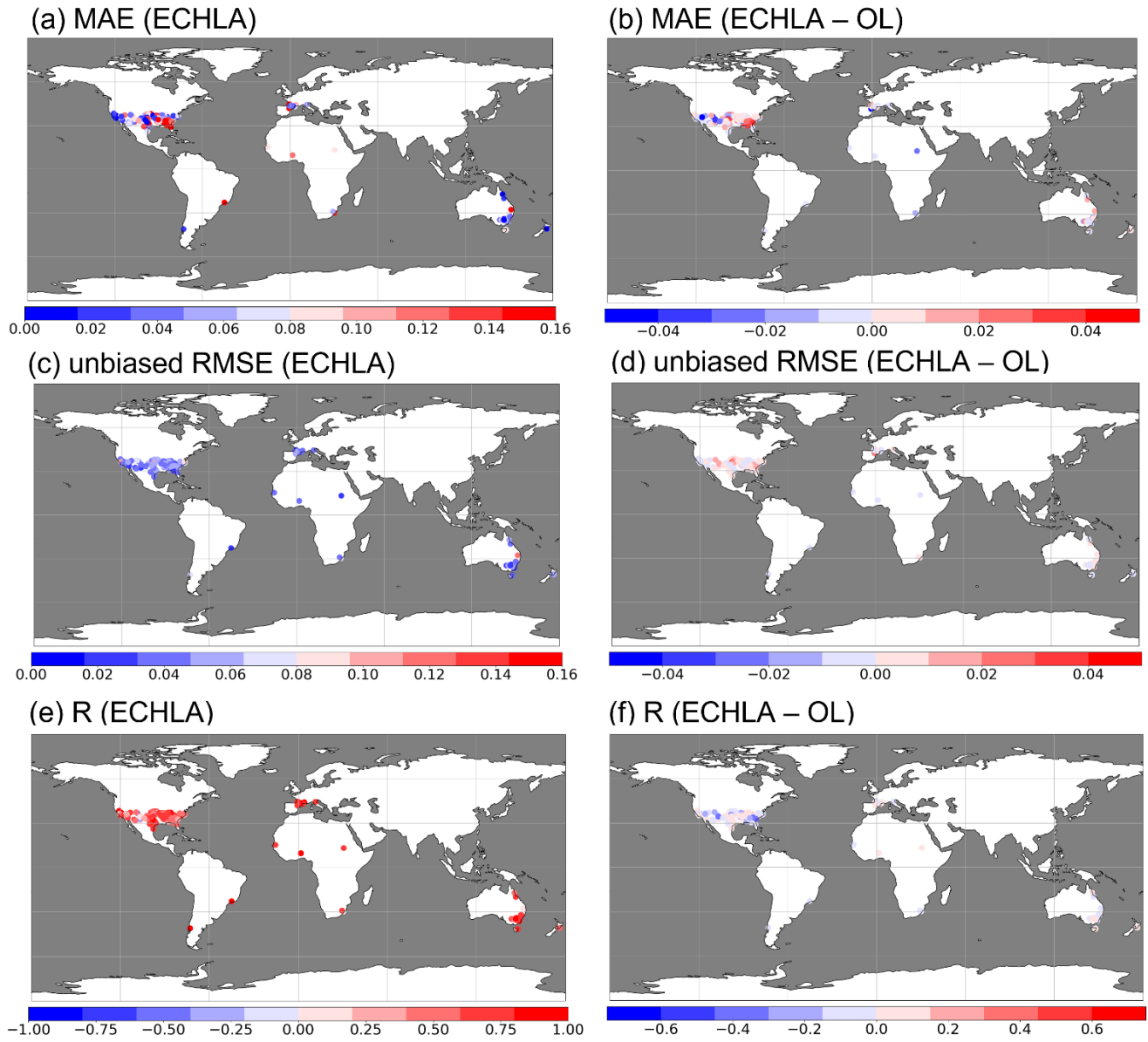

**Figure 7.** Same as Figure 6 but at the depth of 0.05-0.15m.

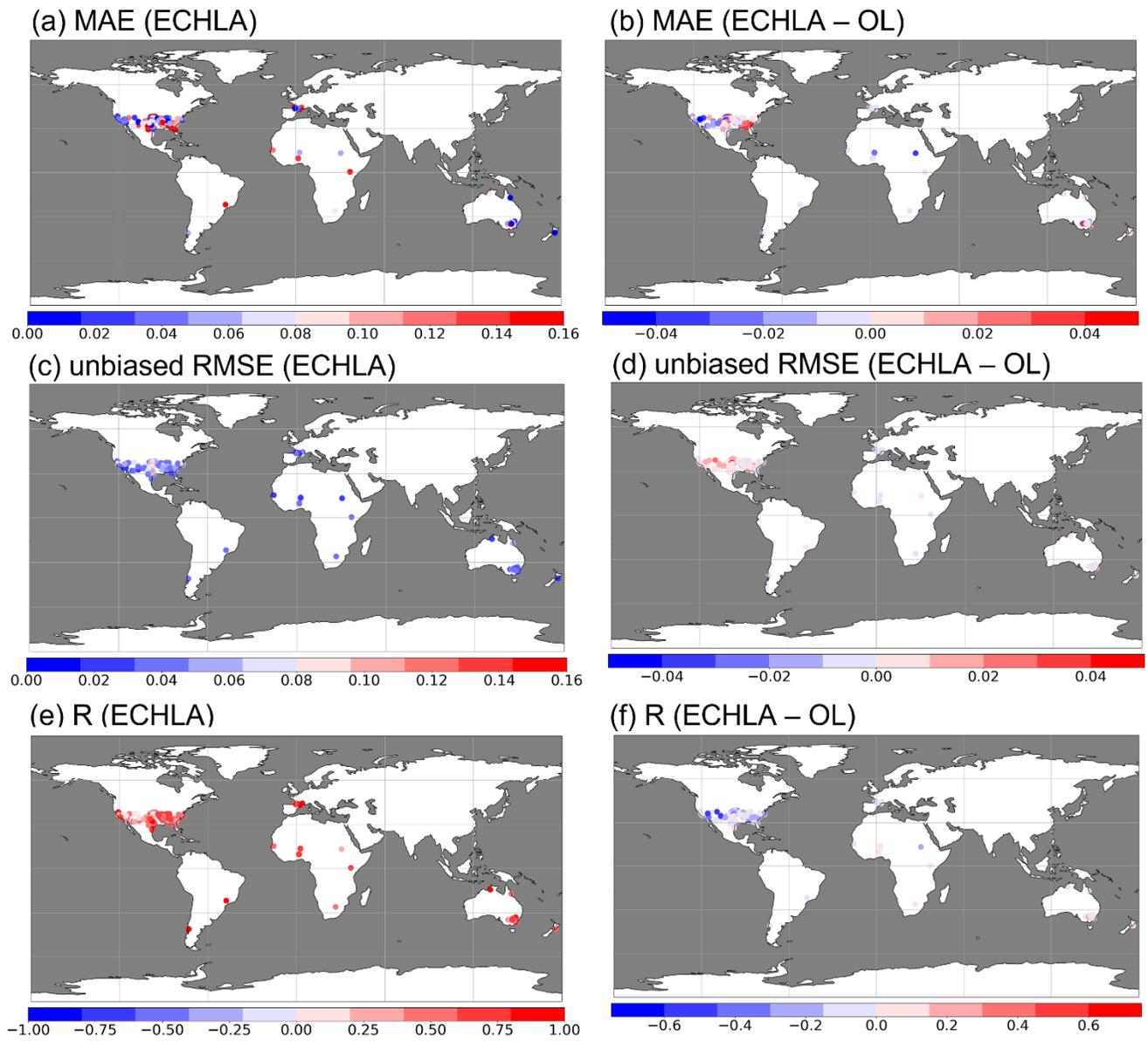

**Figure 8**. Same as Figure 6 but at the depth of 0.15-0.45m.

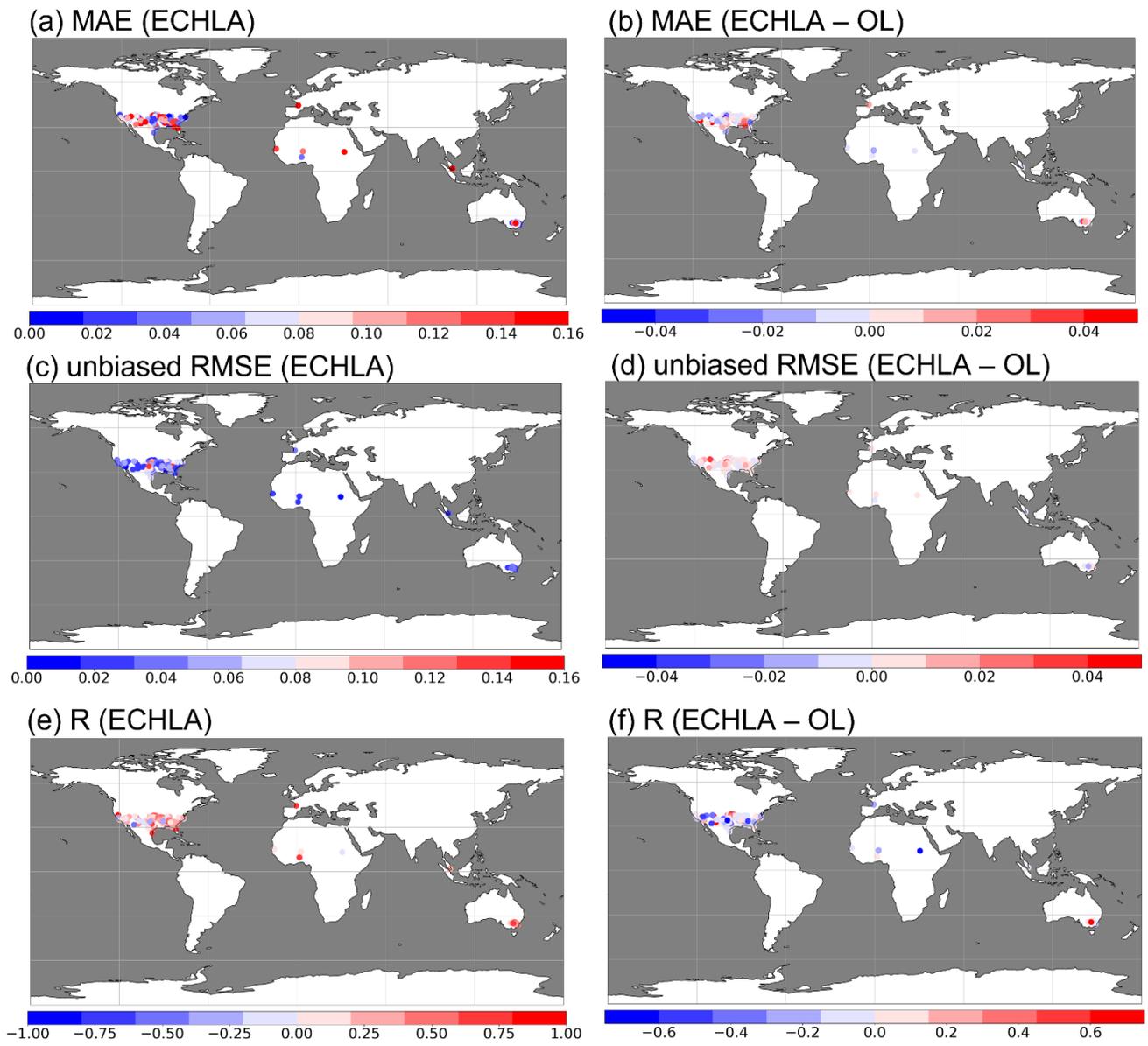

**Figure 9.** Same as Figure 6 but at the depth of 0.45m-1.95m.

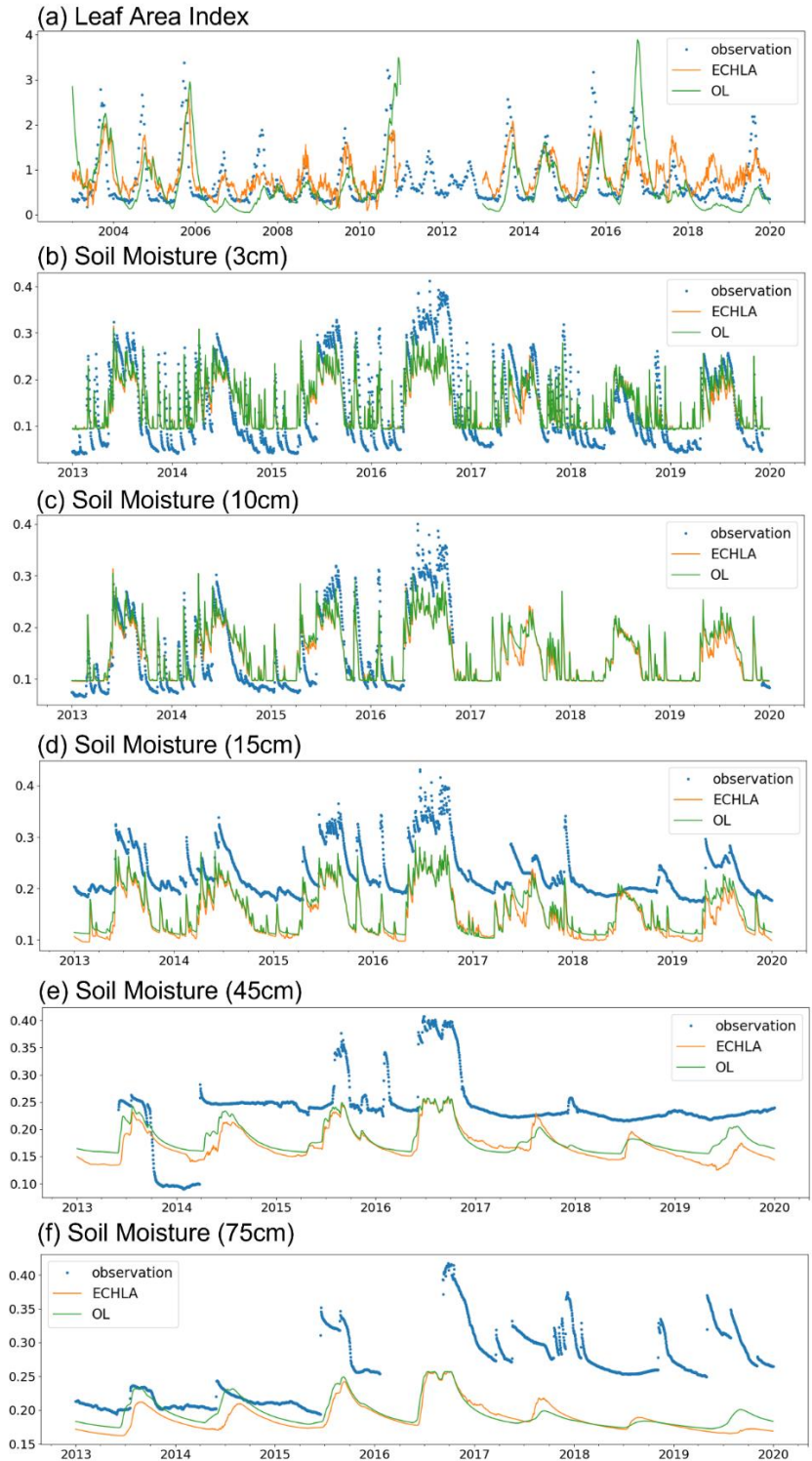

**Figure 10**. Timeseries of (a) LAI, and soil moisture at the depths of (b) 3cm, (b) 10cm, (d) 15cm, (e) 45cm, and (f) 75cm of satellite or in-situ observation (blue dots), ECHLA (orange lines), and the OL experiment (green lines) at the JAXA Yanco calibration/validation site.

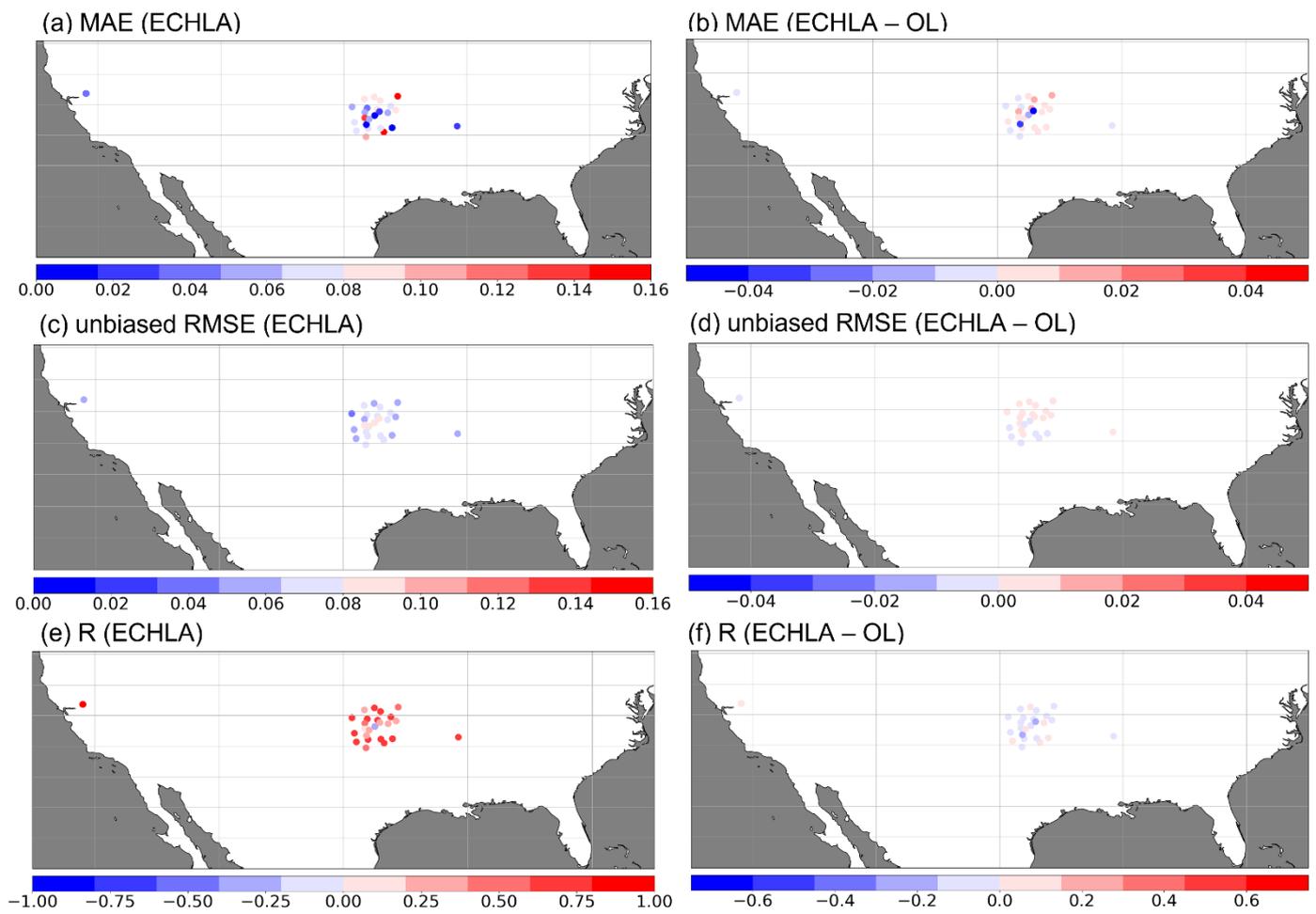

**Figure S1.** Spatial distribution of (a) MAE, (c) ubRMSE, and (e) R of ECHLA soil moisture against ISMN in-situ soil moisture at the depth of 0-0.05m. The differences of (b) MAE, (d) ubRMSE, and (f) R between ECHLA and the OL experiment.

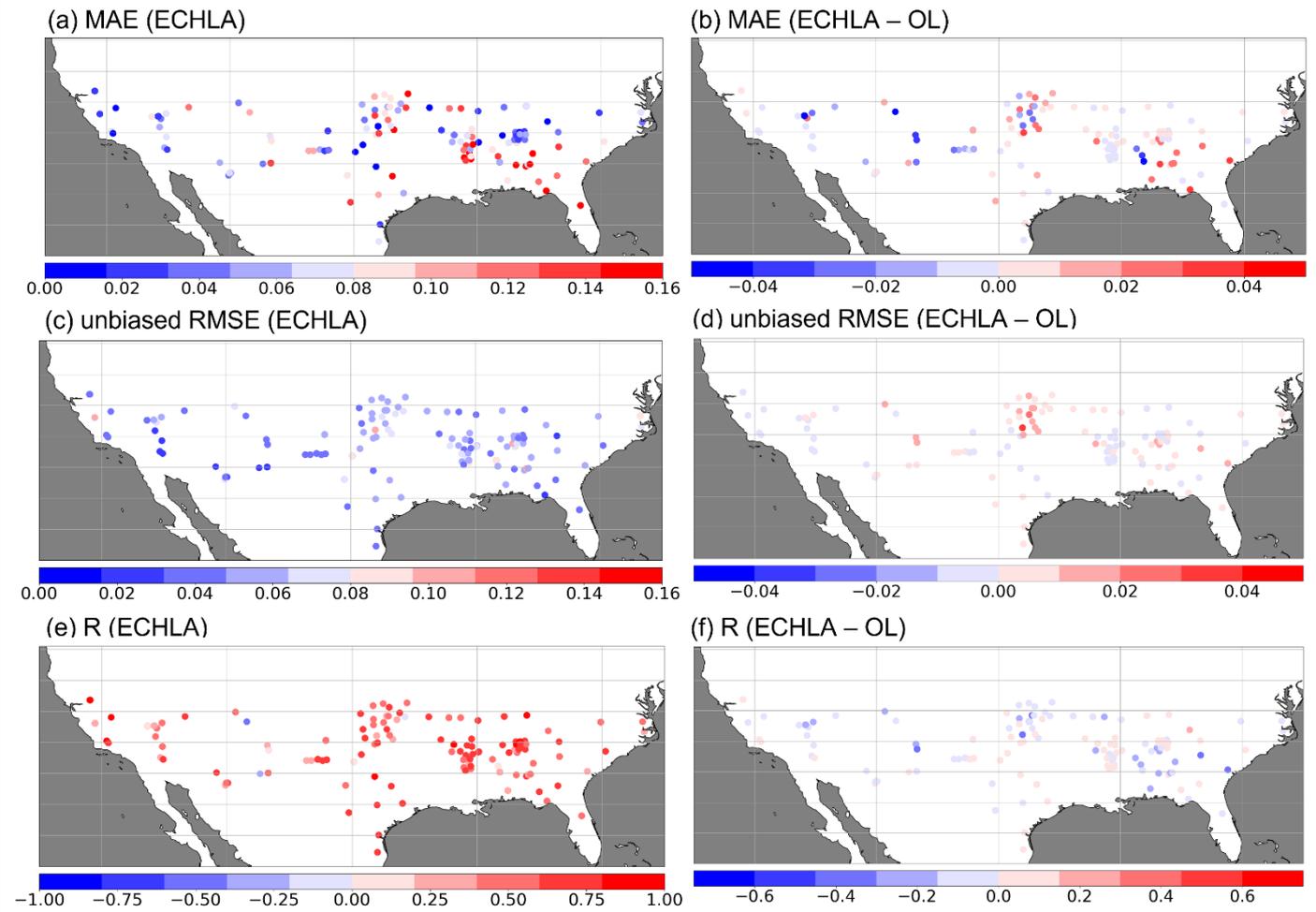

**Figure S2.** Same as Figure S1 but at the depth of 0.05-0.15m

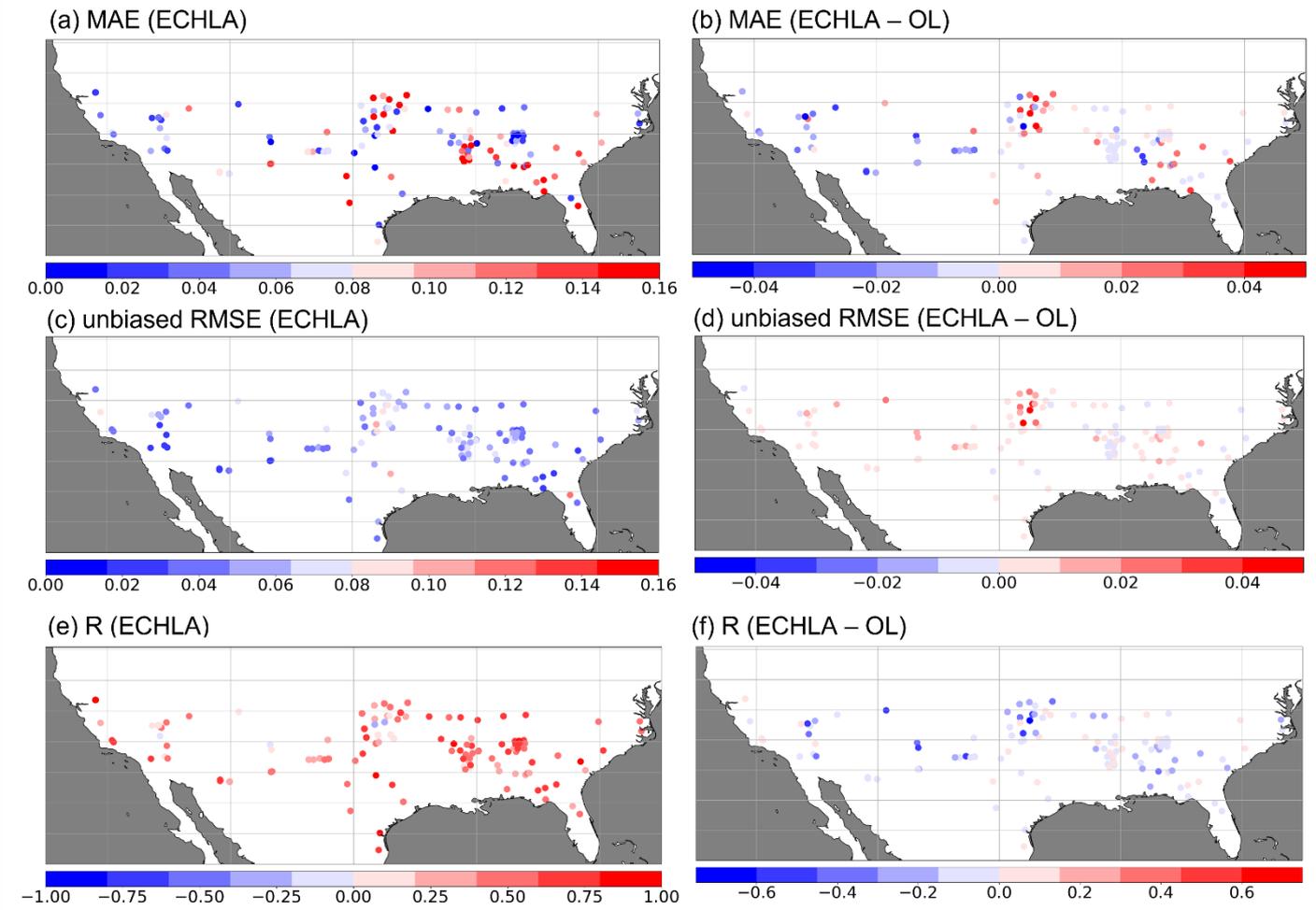

**Figure S3.** Same as Figure S1 but at the depth of 0.15-0.45m

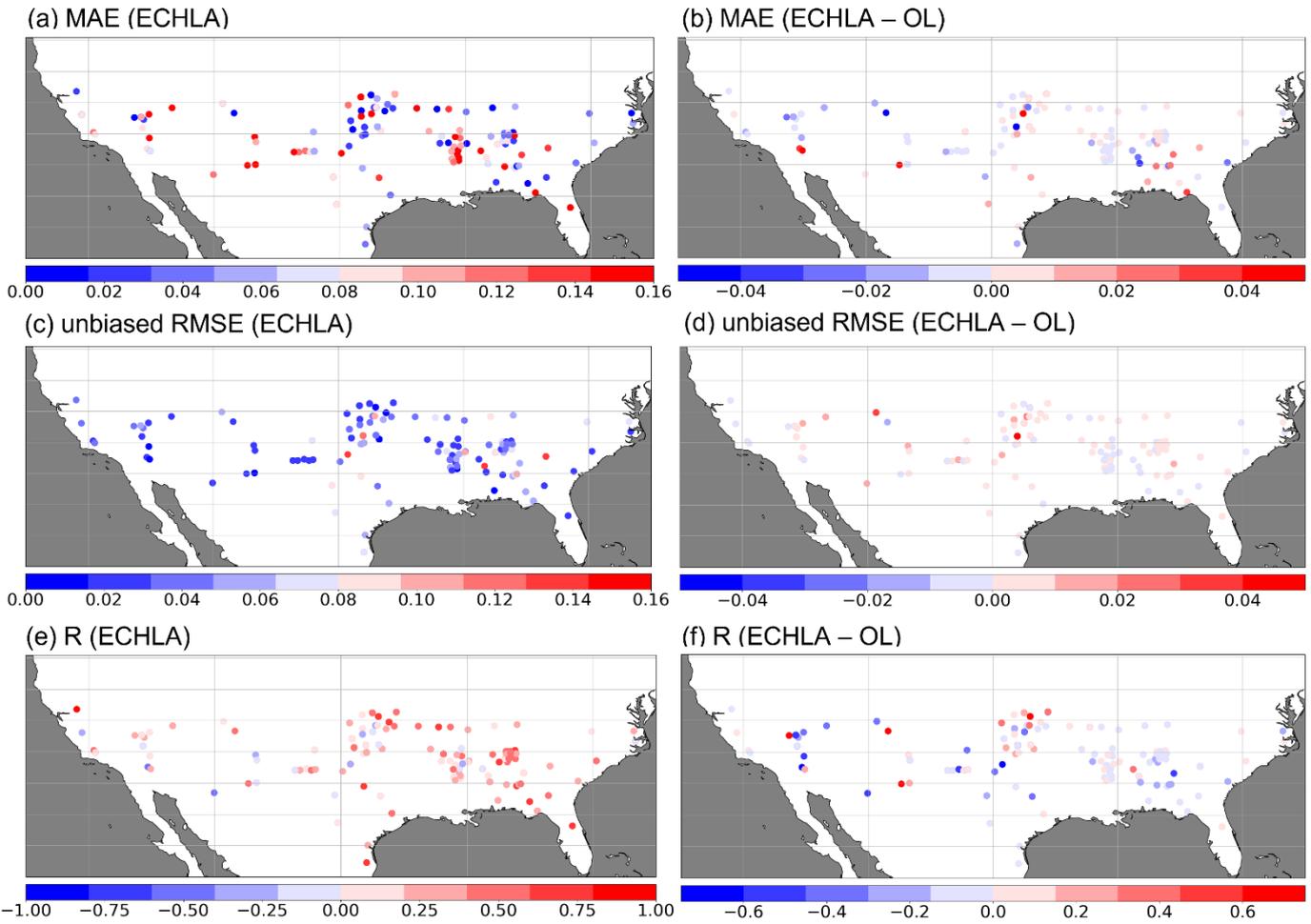

**Figure S4.** Same as Figure S1 but at the depth of 0.45-1.95m

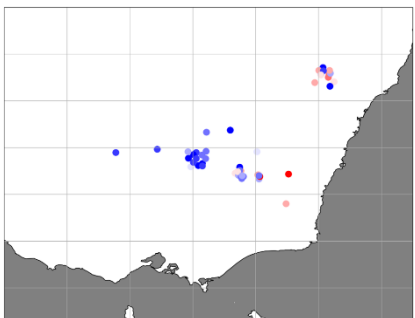
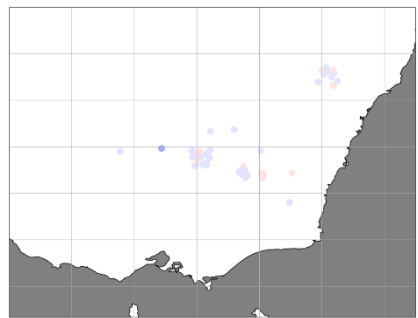
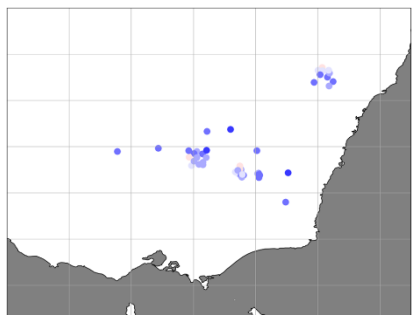
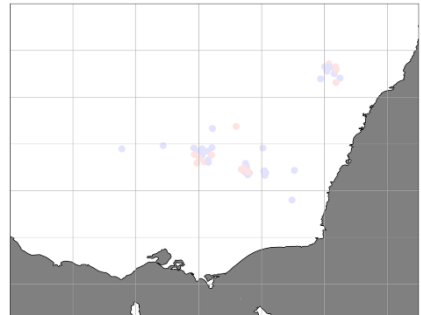
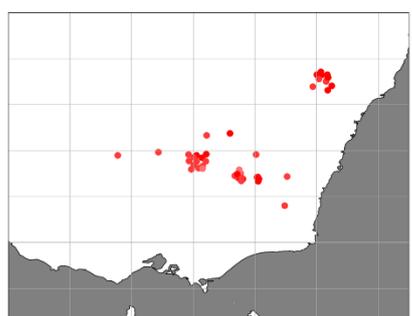
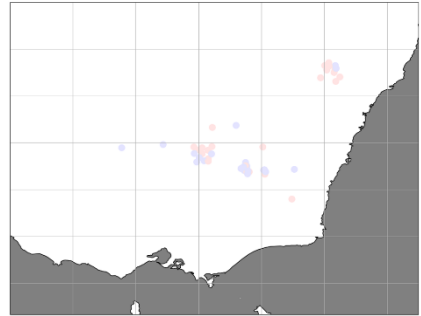

**Figure S5.** Same as Figure S1 but for the observation network in Australia.

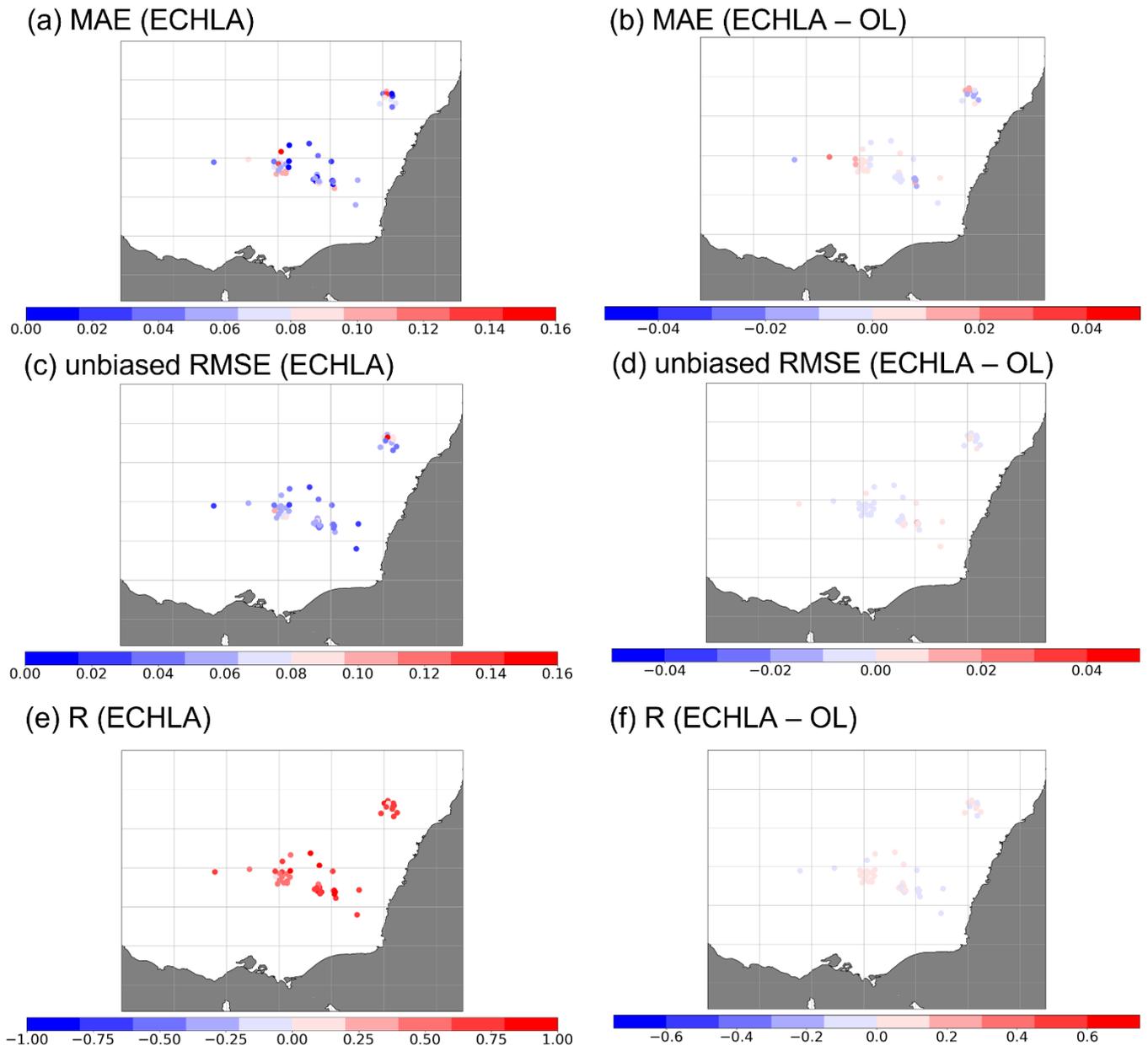

**Figure S6.** Same as Figure S2 but for the observation network in Australia.

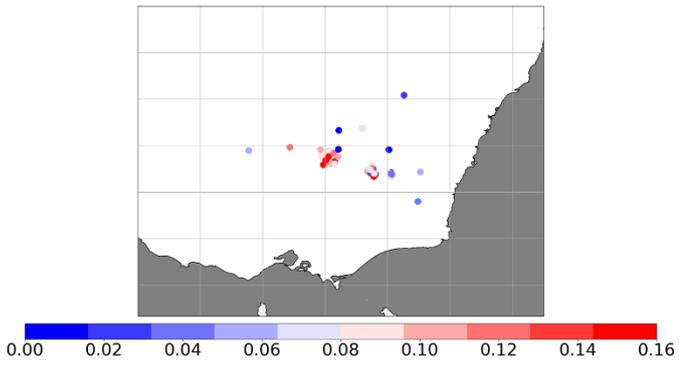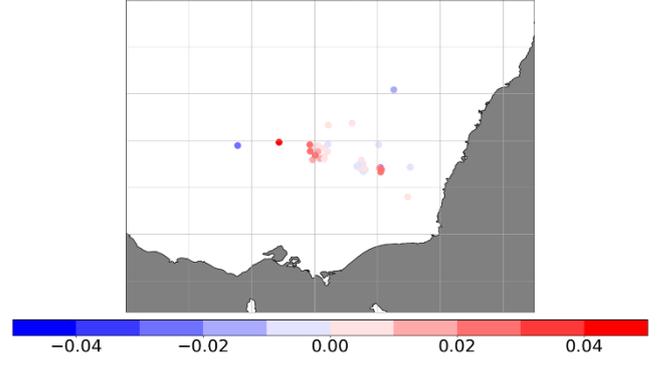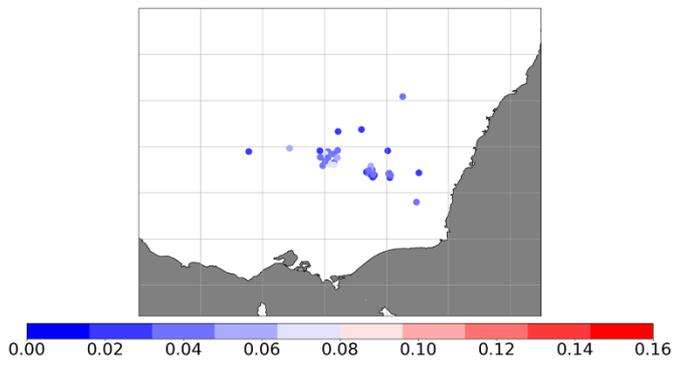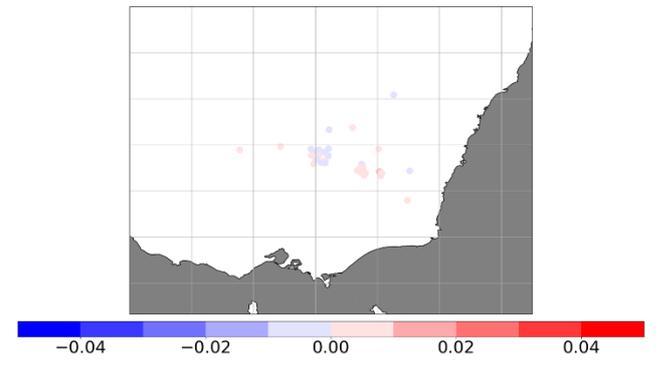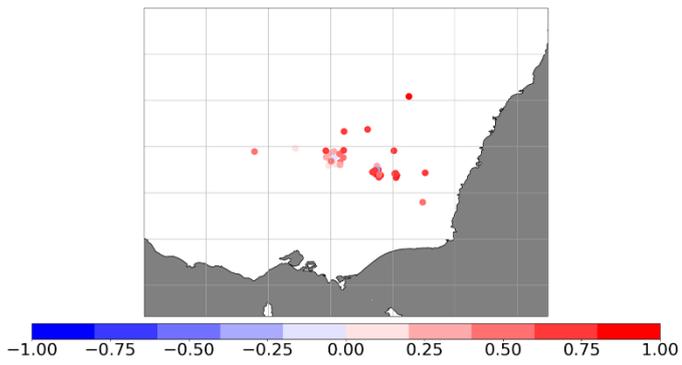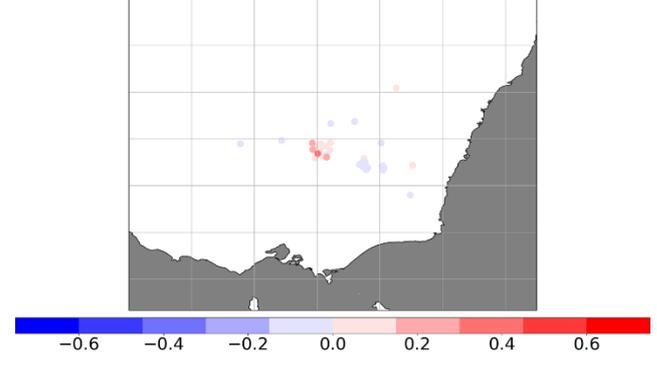

**Figure S7.** Same as Figure S3 but for the observation network in Australia.

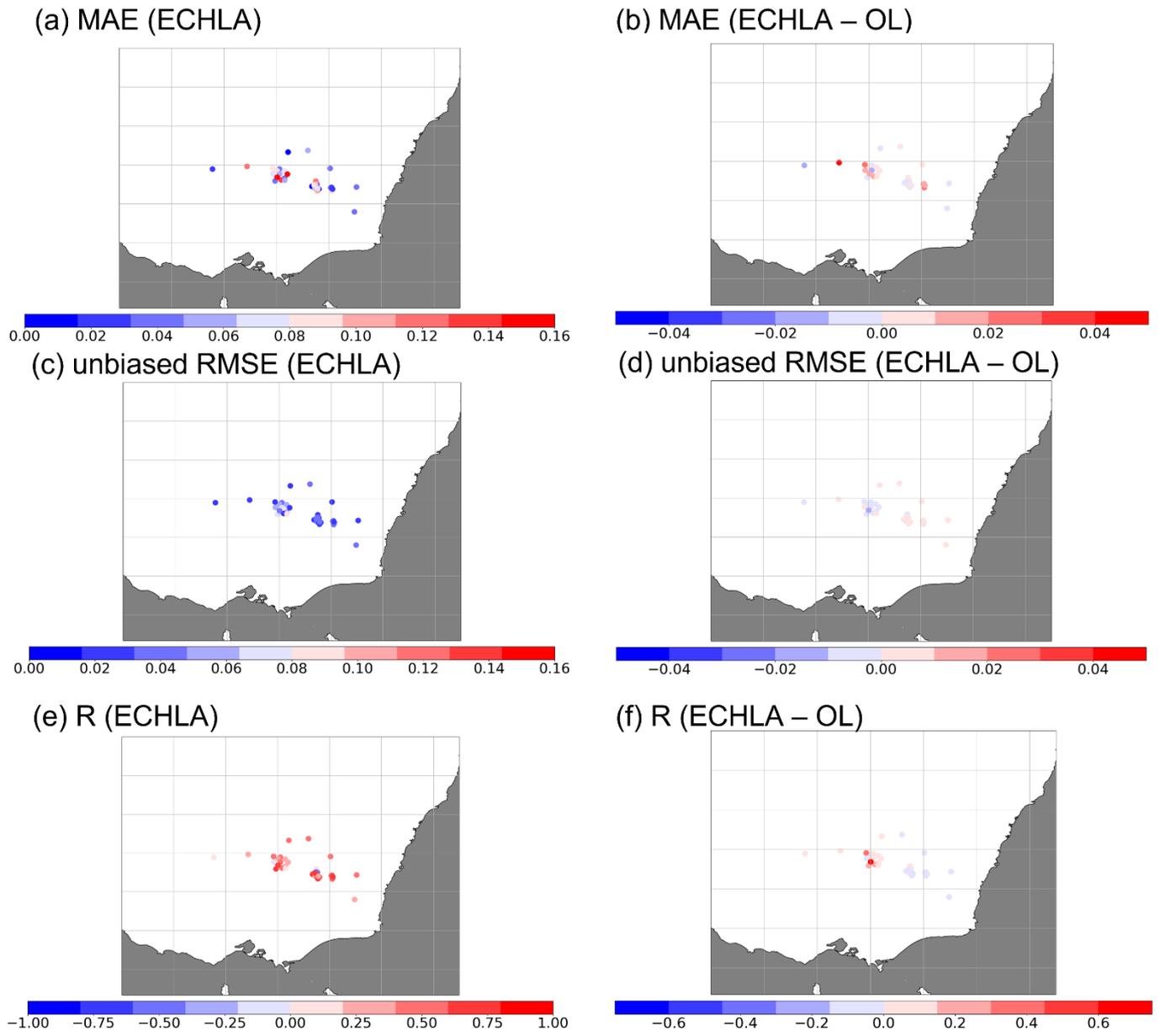

**Figure S8.** Same as Figure S4 but for the observation network in Australia.